%% file: paper.tex
\documentclass[sigconf,nonacm]{acmart}

\usepackage{booktabs}
\usepackage{enumitem}
\usepackage{xspace}
\xspaceaddexceptions{\%}
\usepackage{graphicx}
\usepackage{balance}
\usepackage{url}
\usepackage{tikz}
\usepackage{ifthen}
\usepackage{pmboxdraw}

\input{numbers}

\begin{document}

\title{Packaged, But Not Portable}
\subtitle{Why Conforming to the Agent Plugin Standard Is Rare, and Why
Conforming Would Not Be Enough}

\author{Tezan Sahu}
\email{tezansahu@microsoft.com}
\affiliation{%
  \institution{Microsoft}
  \country{India}
}

\renewcommand{\shortauthors}{Tezan Sahu}

\begin{abstract}
Coding agents are extended by \emph{plugins}: installable bundles that ship
skills, sub-agents, commands, hooks, and tool servers. On \SpecDate, an open
specification---Agent Plugins v1.0.0---standardised how such a bundle is laid
out and described, so that one plugin could run on any agent. We ask the two
questions a practitioner would ask of it: is the ecosystem adopting the
standard, and if a plugin \emph{did} conform, would that be enough to make it
work alongside the other plugins a user has installed?

We answer both by building \CorpusName, a provenance-tracked corpus of \NBundles
plugin bundles across \NRepos repositories, released with its discovery
ledger, scoring code, and analysis. Only \PctConformant\% validate, but the
gap is shallow rather than structural: \PctLoadsFixed\% would load after adding
one missing boilerplate field. The real cost lands elsewhere. \PctLossy\% would
load while the specification obliges the client to \emph{discard} fields their
authors wrote, mostly declarations of what the plugin ships. Moreover, conformance settles nothing for the second question: \PctBundlesExposed\% of
capability-exporting bundles share a name with another plugin, with no
namespace or precedence rule to decide which one answers.

This paper argues the community standardised a packaging \emph{format} when
composition needs a \emph{model}, names the four concepts such a model must
add---qualified capability identity, a declared capability surface, a
precedence rule, and inter-plugin relations---and shows they fit an additive
v1.1 profile of the same specification rather than a competing standard.
Recommendations follow for practitioners packaging extensions today and for the
people evolving the standard, chief among them that conformance must be made
observable before it can become common. The corpus and code are available at
\url{https://github.com/tezansahu/agentpluginzoo}.
\end{abstract}


\maketitle

\section{Introduction}
\label{sec:intro}

Coding agents---GitHub Copilot, Claude Code, OpenAI Codex, Cursor, and their
peers---read and edit a codebase, run commands, and iterate toward a goal
rather than emitting a single completion. Every major one now ships an
extension mechanism, and the unit of extension has converged on a recognisable
shape: a directory containing a \emph{manifest} plus some combination of
\emph{skills} (named instruction files the agent loads when a task matches),
\emph{sub-agents}, \emph{slash commands}, \emph{hooks} that fire on lifecycle
events, and Model Context Protocol (MCP)~\cite{mcp} servers exposing tools. We
call this bundle a \emph{plugin}.

The plugin is now the unit in which agent capability is delivered. Vertical
capability ships as bundles rather than individual tools: published releases
pair domain skills with named connectors in financial services~\cite{anthfin},
life sciences~\cite{anthlife}, and legal work, the last combining twenty-plus
connectors with twelve practice-area plugins~\cite{anthlegal}. Enterprise agent
marketplaces ship their own component vocabularies
alongside~\cite{agentexchange,rovo}. A catalogue is not durability, though: an
earlier assistant plugin ecosystem that reached a little over a thousand
entries was retired roughly a year after launch~\cite{winddown}.

Empirical study of the ecosystem is young and has looked mostly at single
components rather than at bundles or their interaction. Prior work has
censused MCP servers for vulnerabilities~\cite{mcpfirst} and for drift in what
they advertise between versions~\cite{samename}, and mined 42{,}447 skills for
security defects~\cite{skillswild}. A recent survey of the skill
abstraction~\cite{skillsurvey} lists cross-platform portability first among its
open challenges but does not measure it, and package management supplies the
precedent that names are contested resources~\cite{confuguard}---yet we found
no work on capability-name collision between agent extensions.

Two ecosystem facts make composition, rather than packaging, the pressing
problem. First, users install many plugins at once, and hosts already impose
hard limits because of it: one widely-used client caps requests at 128 tools,
reachable by enabling a handful of MCP servers~\cite{toolcap}, and another
budgets the \emph{list} of available skills to 2\% of the context window,
after which it silently omits skills~\cite{skillbudget}. More tools also
measurably degrade selection, with accuracy falling from 93.1\% to 87.1\% as
the offered list grows~\cite{toolcount}. Second, name conflict is not
hypothetical. At least one vendor mandatorily namespaces every plugin skill
``to prevent conflicts when multiple plugins have skills with the same
name''~\cite{namespacing}, an admission that the hazard is real, solved
locally, and not solved portably. A few clients do define a resolution order,
but each defines its own. One documented its tiebreak as ``the first one it
encounters''~\cite{gitlabskills}---traversal order standing in for semantics.

Until recently each agent \emph{vendor}---an organisation such as Anthropic,
GitHub, or OpenAI that builds a client and defines the directory layout it will
load---specified its own convention, so a plugin written for one client had to
be restructured, or duplicated, to run on another. We call a layout
\emph{vendor-specific} when the manifest lives inside a directory named for one
such client, and \emph{vendor-neutral} when it lives at the bundle root the
specification defines. On \SpecDate, Agent Plugins v1.0.0~\cite{spec} was
published to end exactly this fragmentation. This paper investigates what has happened since then, specifically exploring the following questions:

\begin{itemize}[leftmargin=1.9em,itemsep=1pt,topsep=2pt]
\item[\textbf{RQ1}] \textbf{Is the ecosystem converging on the standard?}
\item[\textbf{RQ2}] \textbf{Would conformance be sufficient for quality}---if every plugin conformed tomorrow, would installing several yield predictable behaviour?
\end{itemize}

\paragraph{Contributions.} This paper contributes \CorpusName, a
provenance-tracked corpus of \NBundles bundles from \NRepos repositories across
\NOwners owners, assembled from independent discovery routes with a documented
method for controlling single-route sampling bias (\S\ref{sec:method}). Over it
we report a conformance measurement decomposed into failure modes rather than a
pass/fail bit (\S\ref{sec:conformance}), a measurement of the portability tax
including evidence that \PctDrifted\% of sampled multi-manifest bundles have
diverged (\S\ref{sec:portability}), a temporal analysis
(\S\ref{sec:temporal}), the first quantification of composition hazard in
agent extensibility---capability-name collisions across independently authored
plugins (\S\ref{sec:collisions})---and recommendations for practitioners and
the standard (\S\ref{sec:discussion}).

\section{What the Specification Standardises}
\label{sec:background}

Agent Plugins v1.0.0~\cite{spec} standardises a \emph{package}, not an API.
A plugin is a directory that can be copied, checked into a repository, or
published as an archive, and that any conforming client can open without
out-of-band instructions. The portability claim rests on two decisions: exactly
one \texttt{plugin.json} manifest sits at the package root such that ``no other file
can replace, supplement, or override'' its core fields, and every component
lives at a \emph{fixed, known path} rather than being listed anywhere. A client
does not ask the manifest what a plugin contains. It walks
\texttt{skills/*/SKILL.md} and reads \texttt{mcp.json}. What matters here is
that the two are disconnected (Fig.~\ref{fig:layout}): the manifest carries
identity and provenance
only---\texttt{\$schema}, \texttt{name}, \texttt{version}, \texttt{description},
\texttt{author}, \texttt{license}, \texttt{repository}---and says nothing about
what the plugin does.

The package is deliberately narrow. Version~1.0.0 standardises exactly
\textbf{two component types}---skills and MCP servers---and states that other
types are outside the v1 format and \emph{do not affect conformance}. Commands,
hooks, sub-agents, rules, and language servers are deferred as too
client-specific to fix yet. Vendors are not left without a home: a top-level
reverse-domain directory (\texttt{com.example.client/}) is sanctioned for
client-specific files, as is an \texttt{extensions} object inside the manifest
keyed the same way. But the specification ``assigns no portable discovery,
validation, loading, or failure semantics'' to either.

Three further properties of the schema~\cite{schema} drive our results.

\begin{figure}[t]
\centering
\scriptsize\ttfamily
\begin{minipage}[t]{0.44\columnwidth}
my-plugin/\\
├──~plugin.json\\
├──~skills/\\
│~~~~└──~code-review/\\
│~~~~~~~~~└──~SKILL.md\\
├──~mcp.json\\
└──~com.example.client/
\end{minipage}\hfill
\begin{minipage}[t]{0.52\columnwidth}
\{\\
\hspace*{0.6em}"\$schema": "...",\\
\hspace*{0.6em}"name": "my-plugin",\\
\hspace*{0.6em}"version": "1.0.0",\\
\hspace*{0.6em}"description": "...",\\
\hspace*{0.6em}"author": \{"name": "..."\},\\
\hspace*{0.6em}"license": "MIT"\\
\}
\end{minipage}
\caption{Layout (left) and manifest (right) are disconnected.}
\label{fig:layout}
\end{figure}

\begin{itemize}[leftmargin=1.2em,itemsep=1pt,topsep=2pt]
\item It is \textbf{closed}: \texttt{additionalProperties} is \texttt{false},
      so any field the schema does not name is a violation.
\item It \textbf{requires \texttt{\$schema}} alongside \texttt{name}. A
      manifest omitting it cannot validate, however well-formed it otherwise
      is.
\item Violations are \textbf{not equal}. An unknown top-level field, or a
      non-object \texttt{extensions}, is \textit{non-fatal}: a client ``MUST report and
      ignore each unknown field and MUST continue loading the plugin.'' Any
      other violation is \textit{fatal}---the client ``MUST reject the plugin and MUST
      NOT discover or execute any of its components.''
\end{itemize}

That last asymmetry is the hinge of \S\ref{sec:conformance}. A plugin can be
non-conformant in two categorically different ways: it can fail to load at all,
or it can load perfectly while the specification obliges the client to throw
away what its author wrote. A generalized validator counts both as invalid. However, only one is a portability problem.

Equally important is what v1.0.0 does \emph{not} define. There is no
namespacing for capability names, no precedence rule when two installed plugins
expose the same capability, and no dependency or compatibility relation between
plugins. Plugin names form a flat namespace of 1--64 lowercase characters with
no owner scoping. Even \texttt{version} is a bare string: semantic
versioning~\cite{semver} is recommended but not required, and a client ``MUST
NOT'' reject a plugin for violating it, so hosts cannot generally order two
releases. The roadmap acknowledges part of this, noting that ``plugins
currently cannot declare dependencies on other plugins''~\cite{future}, but
precedence and name collision appear nowhere. The specification describes a
plugin in isolation, and is silent on a set of plugins, which is the only
configuration users actually run.

\section{\CorpusName: A Plugin Bundle Corpus}
\label{sec:method}

\CorpusName holds \NBundles plugin bundles, \NSkillsStandalone standalone
skills, and \NMcpStandalone standalone MCP configurations across \NRepos public
GitHub repositories, with every artifact tagged by the query that surfaced it.
Throughout, a \emph{repository} is one public GitHub repository and an
\emph{artifact} is one thing we found inside it. It is not dominated by
mega-repositories: \PctSingleBundleRepos\% of contributing repositories hold
exactly one bundle and the largest accounts for \PctTopRepoShare\% of bundles.
Because code search skips forks, they are effectively absent from the corpus.

\subsection{How it is built}

Collection runs in three phases. \textbf{Discovery} nominates candidate
repositories: we issue \emph{queries} against public GitHub~\cite{ghsearch}
along several independent routes---code search for manifests at each known
vendor-specific location, for manifests declaring the v1.0.0 \texttt{\$schema},
and for marketplace index files, plus repository search on descriptive terms.
Each route is a different guess about where plugins live, with a different
bias. \textbf{Expansion} turns a repository into artifacts: for every nominated
repository we make one recursive Git tree call, returning the full path listing
of the default branch in a single response, then walk it offline to identify
every directory that looks like a bundle---whether or not any query would have
matched it. \textbf{Extraction} fetches the manifest and capability files for
each bundle, parses them, and records the result alongside the route that first
surfaced it.

Table~\ref{tab:ledger} reports each route's yield, and the released ledger
records every query verbatim, including routes that did not contribute---
notably the non-GitHub MCP server registry---so the corpus is auditable for
what it omits as well as what it contains.

\begin{table}[t]
\caption{Discovery routes and their yield.}
\label{tab:ledger}
\small
\begin{tabular}{@{}lrrr@{}}
\toprule
\textbf{Route} & \textbf{Queries} & \textbf{Repos} & \textbf{Artifacts} \\
\midrule
\input{ledger}
\\
\bottomrule
\end{tabular}
\end{table}

\subsection{Safeguards against self-measurement}
\label{sec:bias}

A code-search query can only return files matching what it asks for. If we ask
GitHub for files at \texttt{.some-vendor/plugin.json}, every result sits in
that vendor's directory by construction, so reporting how many plugins use that
layout would measure our own query rather than the ecosystem. An early pass
searching one vendor's manifest path alone found 89\% of bundles carrying that
vendor's directory. We apply four safeguards.

\begin{itemize}[leftmargin=1.2em,itemsep=1pt,topsep=2pt]
\item \textbf{Expansion, not search, does the work.} Code search is capped at
1{,}000 results per query and saturates almost immediately, while one recursive
tree call enumerates every bundle in a repository under a far more generous
quota. Discovery need only nominate a repository once, and expansion then finds
bundles in layouts we never searched for, decoupling what we find from what we
asked. It produced the majority of artifacts.
\item \textbf{Ordering.} Vendor-neutral routes run before vendor-specific ones,
so the corpus does not start out skewed toward any single layout.
\item \textbf{Partitioning.} On routes that exceeded the result ceiling we
recursively partition each query on the \texttt{size:} qualifier until every
partition falls under it. Where one remains truncated, the ledger records the
reported total alongside the number retrieved.
\item \textbf{Recomputation.} Every artifact records which queries surfaced it,
so any result that might depend on a particular query can be recomputed over
just the sub-corpus that was \emph{not} found by it. We report such results
both ways in \S\ref{sec:portability}.
\end{itemize}

\section{Why Conformance Fails}
\label{sec:conformance}

\begin{figure}[t]
\centering
\includegraphics[width=0.93\columnwidth]{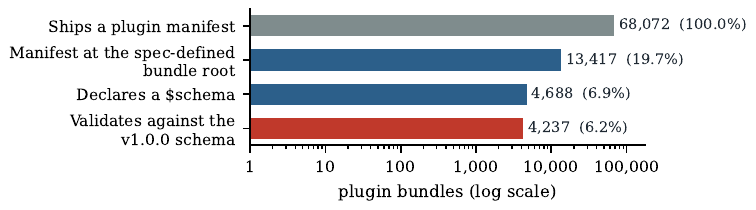}
\caption{Conformance ladder over all \NBundles bundles.}
\label{fig:funnel}
\end{figure}

We scored every bundle against the official v1.0.0 manifest
schema~\cite{schema} using a standard Draft 2020-12
validator~\cite{jsonschema}. Rather than collapse the result to pass/fail, we
decompose it into the ladder, as shown in Figure~\ref{fig:funnel}.

Only \NLocated bundles (\PctLocated\%) place a manifest at the
specification's bundle root at all. The rest keep it inside a vendor-specific
directory. Of those correctly located, a minority declare the required
\texttt{\$schema}. End to end, \NConformant bundles---\PctConformant\% of the
corpus---validate against the standard. A low number would be a dull finding,
as ecosystems adopt standards slowly. Why they fail is the interesting result.

\paragraph{Not all failures are equal.}
Validation is binary, but the specification is not. As \S\ref{sec:background}
noted, an unknown top-level field is explicitly non-fatal, while any other
violation obliges the client to reject the plugin outright. We re-scored every parseable manifest applying those two exemptions, to ask a
sharper question than ``does it validate?'': \emph{would a conforming client
load it?}

Based on our findings, it would load \PctLoads\%, while \PctRejected\% would be rejected outright, but the
cause is almost entirely a single missing line: granting every manifest a
correct \texttt{\$schema} declaration and changing nothing else,
\PctLoadsFixed\% load. Only \NDeepFatal bundles (\PctDeepFatal\%) fail for a
deeper reason, two-thirds of those being name-constraint violations. Every
conformance rate elsewhere is computed strictly against the published schema.

This is both encouraging and troubling. Structurally the ecosystem
may \emph{already} be compatible with the standard, since non-conformance is
overwhelmingly a boilerplate field nobody has heard of rather than incompatible
design. But among manifests that would load after that one edit, \PctLossy\%
(\NLossy bundles) carry at least one top-level field the specification then
requires the client to \emph{discard}. Such a plugin succeeds while the host throws away what the author wrote, a worse failure mode than rejection because nothing reports it.

\begin{figure}[t]
\centering
\includegraphics[width=0.93\columnwidth]{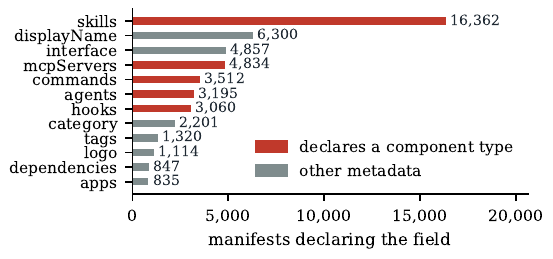}
\caption{Top-level fields the closed v1.0.0 schema forbids.}
\label{fig:forbidden}
\end{figure}

Figure~\ref{fig:forbidden} ranks the fields present in real manifests that the
closed schema rejects, led by \texttt{skills}, \texttt{mcpServers},
\texttt{commands}, \texttt{hooks}, and \texttt{agents}. The first two are the
most telling: they are precisely the component types v1.0.0 \emph{does}
standardise, and the specification's answer to both is the package layout. A
client finds skills by walking \texttt{skills/} and tool servers by reading
\texttt{.mcp.json}, so the manifest has nothing to say about them. Authors
writing these fields are not asking for a feature. They are declaring, in the
one file a reader opens first, what the directory structure already
implies---convention-based discovery is evidently not self-evident to the
people doing the packaging. \PctDeclareComponents\% of bundles do this.

The remaining entries ask for something the format genuinely lacks. Commands,
hooks, and sub-agents were deferred as too client-specific
(\S\ref{sec:background}), so authors writing them describe capabilities their
plugin really ships, for which no vocabulary exists. Two rarer fields matter: \texttt{dependencies}, expressing that one plugin requires another, and
\texttt{userConfig}, declaring configuration the host should collect before
activation. Both are composition concerns with nowhere to live in v1.0.0.

\paragraph{The escape hatch nobody found.}
The specification does sanction a place for unstandardised data: the
\texttt{extensions} object, keyed by reverse-domain namespace. Only
\PctExtensions\% of manifests use it, across \NExtNamespaces namespaces,
against the \PctAnyForbidden\% that write a forbidden field at the top level
instead. The mechanism exists, is documented, and would have made many of these
manifests valid. Authors did not find it, which is a discoverability failure
and the cheapest finding to act on.

The typical non-conformant plugin is therefore not a careless one. It is a plugin
whose author wanted to state what it \emph{does} and found the manifest has no
way to say it. The cost is that a host cannot learn what a plugin exposes
without downloading and walking it, so conflicts cannot be detected before
installation.

\section{The Portability Tax}
\label{sec:portability}

If a vendor-neutral format exists, the portable move is to ship one manifest at
the bundle root. However, most authors do not: only \PctLocated\% place a manifest there. The dominant pattern is a manifest inside a single vendor's directory,
so the plugin runs on one agent and must be restructured for any other, while a
visible minority ship the same manifest in several vendor directories, paying
for portability by duplication. Location is precisely the property our
vendor-path queries could have biased (\S\ref{sec:bias}), so we recomputed it
over the sub-corpus assembled \emph{without} any vendor-path query
(\NNeutralBundles bundles) and the shape holds.

The clients themselves document the tax. At least two now describe supporting
the portable root manifest \emph{and} a client-native one in the same
repository as a compatibility fallback~\cite{codexpkg,cursorplugins}, and one
further documents that a path variable the specification defines is expanded
only under a client-prefixed name~\cite{cursorplugins}. Dual-manifest authoring
is the documented path, not a misunderstanding, and maintainers describe the
same situation from the other side: skills shipped under per-client
directories, with no registry to distribute them from~\cite{talkskillsmcp}.

The cost is a \emph{portability tax}: to support $k$ agents an author maintains
$k$ copies of the same metadata, or picks one agent and accepts lock-in. It
shows up directly in conformance. Bundles adopting no vendor directory conform
at \ConfNeutralOnly\% (n=\NNeutralOnly), those inside exactly one---the
lock-in case---at \ConfOneVendor\% (n=\NOneVendor), and those spanning two or
more recover to \ConfMultiVendor\% (n=\NMultiVendor). This is likely because an author supporting several agents has reason to put something at the neutral root too. Thus, vendor commitment correlates with abandoning of the standard.

\paragraph{Duplicated manifests do not stay in sync.}
Shipping $k$ copies works only if the copies agree. Within \CorpusName, \NMultiManifest bundles
(\PctMultiManifest\%) ship two or more manifest files. We drew a random sample
of \NDriftSampled and fetched every copy. \PctDrifted\% had copies that were
not semantically identical, and \PctDriftName\% disagree on \texttt{name}, so
the bundle announces itself differently depending on which agent loads it.
Prior work found similar silent drift in one server \emph{across
time}~\cite{samename}. This is drift across one plugin's own copies at one
instant. On the specification's own terms duplication is not a workaround at
all: v1.0.0 admits ``exactly one portable manifest per plugin.'' The standard
removed the technical reason for this tax two months before our snapshot, yet
\S\ref{sec:temporal} shows the behaviour persisting regardless.

\section{Conformance Is Rising, From a Very Low Base}
\label{sec:temporal}

\begin{figure}[t]
\centering
\includegraphics[width=0.93\columnwidth]{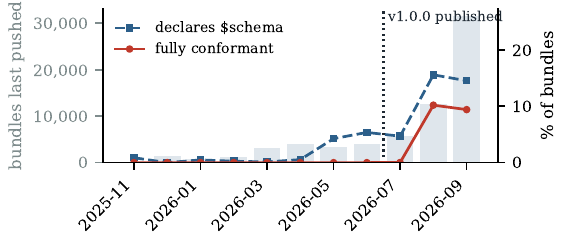}
\caption{Conformance against repository last-push month.}
\label{fig:temporal}
\end{figure}

\paragraph{Establishing the date.} The specification repository has no releases
or tags, so we determined the date from commit history. It was created in April
2026 carrying an early draft under a different project name, lay dormant three
months, then a signed commit titled ``Publish Agent Plugins Specification
1.0.0'' landed on \SpecDate. We use that commit, since repository creation
would contaminate the post-publication bucket with three months of earlier
work.

Splitting \CorpusName at that date by repository creation, conformance is
\ConfPre\% before (n=\NPre) and \ConfPost\% after (n=\NPost), a better than
sixfold increase. Creation date is a weak proxy for when a manifest was
authored, so we also take cuts that do not depend on it. \NLive bundles live in
repositories \emph{pushed to} after publication, whose maintainers were
demonstrably active in the post-standard world, and conformance among them is
\ConfLive\%. Narrowing to repositories touched in the final 30 days of our
window (n=\NRecent) it reaches \ConfRecent\%, with \SchemaRecent\% declaring a
\texttt{\$schema}. Monotonicity across three independent slices is itself
evidence that adoption is accelerating.

\paragraph{The number that matters is the other one.} It would be easy to read
a sixfold rise as success. But an overall \ConfRecent\% conformance among the most actively maintained projects means roughly nine in ten freshly-touched bundles
still do not validate, two months after a standard existed to validate against.
More importantly, the trend is confined to the part of the problem the format
addresses. Nothing in the specification moved on naming, precedence, or
dependency, so nothing in the data moved either.

\section{Plugins Collide, and Nothing Says Who Wins}
\label{sec:collisions}

Conformance concerns one plugin. But practically, users install many, and the
hosts they run on already show the strain through the tool caps and skill
budgets noted in \S\ref{sec:intro}. The hazard is understood---one vendor
namespaces every plugin skill by force to prevent it~\cite{namespacing}---and
simply not addressed by the standard. We therefore extracted the \emph{names}
of every capability each bundle in \CorpusName exports---skills, sub-agents,
commands, hooks---yielding \NCaps records.

\begin{figure}[t]
\centering
\includegraphics[width=0.93\columnwidth]{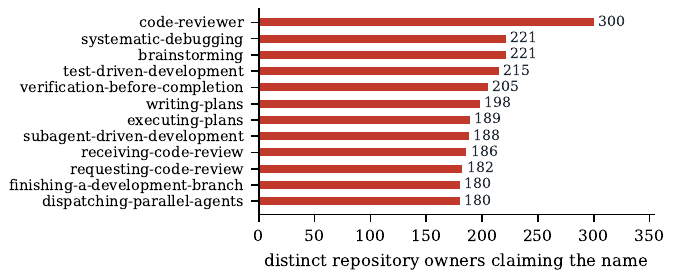}
\caption{The most-contested capability names, by distinct owner.}
\label{fig:contested}
\end{figure}

Skill names are heavily reused: \NSkillNames distinct names across the corpus,
of which \PctSkillNamesShared\% are claimed by more than one plugin. Because
contested names are also the popular ones, \PctSkillInstContested\% of
individual skills carry a name that some other plugin also uses. Aggregating
across kinds, \PctBundlesExposed\% of the \NCapBundles bundles that export a
named capability share at least one name with another plugin.

Figure~\ref{fig:contested} names the worst cases. The most contested name,
\texttt{\TopNameOne}, is claimed by \TopNameOneOwners distinct repository
owners, and the runner-up, \texttt{\TopNameTwo}, by \TopNameTwoOwners. Others
in the head of the distribution too refer to \textit{debugging} and \textit{code review}. These are
exactly the capabilities a user installs a plugin to get: generic, task-shaped
names for common jobs. This is not accidental collision in a large namespace,
but convergence on the obvious name.

\paragraph{Is this just vendoring?} A name repeated across many bundles inside
one organisation is duplication, not a conflict a user could hit, so we count
it separately everywhere. The most-duplicated skill name appears in
\MaxSkillClaims bundles but from only a handful of owners, making it a
vendoring artifact. Requiring the rival claim to come from a \emph{different}
owner, exposure falls only from \PctBundlesExposed\% to
\PctBundlesExposedXOwner\%. Collisions are overwhelmingly between independent
publishers, precisely the case no host can resolve by policy.

Now consider the user who installs two such plugins. Which skill answers? Is
the other shadowed, merged, or an error? Does the answer change with install
order? Agent Plugins v1.0.0 does not say, because it never introduces the
concepts needed to say it. Resolution is left to each host, so the same two
plugins behave differently on different agents---exactly the cross-agent
divergence that the specification was written to end.

\section{Discussion}
\label{sec:discussion}

\subsection{Answering the two questions}

The curation and analysis of \CorpusName helps us address both the questions that we decided to investigate:

\paragraph{RQ1: is the ecosystem converging?} Yes, but slowly and from a very
low base. \PctConformant\% of the corpus conforms, rising to \ConfRecent\%
among actively maintained projects, and the rate more than sixfolds across the
publication date. The direction is unambiguous, the level marginal. The gap is
also \emph{shallow}: \PctLoadsFixed\% would load after one added field, making
this a gap in awareness and tooling, not compatibility.

\paragraph{RQ2: would conformance be sufficient?} No. Every plugin in our corpus
could validate tomorrow and \PctBundlesExposed\% of capability-exporting
bundles would still sit on a name another plugin claims, with nothing in the
standard to decide which answers.

The specification can be satisfied in full and still leave this open, because
it describes each plugin on its own. What it standardises is a \emph{format}:
the shape of one artifact, checkable on a plugin sitting alone on disk. What
collisions demand is a \emph{model}: rules covering plugins in relation, naming
what a capability \emph{is}, how it is identified, what counts as a clash, and
what a host must do when one occurs. A format is satisfied by a file, a model
by an \emph{installation}---so the question only arises once a second plugin is
present, which is the only configuration users actually run.

\subsection{What a composition model must name}

Authors ask for expressiveness the specification forbids
(\S\ref{sec:conformance}), duplicated metadata drifts across vendor copies
(\S\ref{sec:portability}), and independently authored plugins claim the same
names with no rule to resolve them (\S\ref{sec:collisions}). A standard
enumerating component \emph{types} addresses none of these. However, a model defining the following concepts would:

\textbf{1. Qualified capability identity.} A capability should be addressable
as \texttt{plugin:capability}, with the bare name as a user-facing alias---the
minimum needed to state a conflict, let alone resolve it. One vendor already
does this unilaterally~\cite{namespacing}, which shows it works and that
per-vendor namespacing is not portability.

\textbf{2. A declared capability surface.} Let the manifest say what the plugin
exposes, the very thing \PctDeclareComponents\% of authors already attempt.
Today a host must download and walk a bundle to learn its contents, putting
registries, search, and pre-install checks out of reach. Declaration moves that
information into the file read first, leaving convention-based discovery the
default.

\textbf{3. A precedence and conflict rule.} Hosts need a defined outcome for a
name claimed twice: a deterministic winner, a user disambiguation step, or a
hard error. Any defined answer beats today's, which is per-host and invisible
to the user.

\textbf{4. Composition relations between plugins.} Dependency, incompatibility,
and version constraints already appear in the wild as invented manifest fields,
and the roadmap acknowledges the gap~\cite{future}. Dependency solving is
NP-complete in any non-trivial component model~\cite{depsolving}, so this needs
design rather than improvisation.

\paragraph{Why this has not happened.} Concepts 1 and 3 have been put to the
working group and declined. An almost identical proposal for portable component
identities drew the response that this is ``an intentional omission'', because
dictating collision logic ``would be too much of an imposition'' on client
implementers~\cite{identdisc}. That is a judgement about cost made without a
measurement of the cost on the other side, which is what \CorpusName supplies.
Delegating resolution to clients does not make the problem client-sized. It
makes resolution unportable, the one property the specification exists to
supply. Namespacing is meanwhile being added at the \emph{intra}-plugin scope,
where collisions were never the problem~\cite{nsskills}. The venue is open,
though: a 1.1.0 working draft~\cite{draft11}, a registry index~\cite{regdisc},
and a dependency field~\cite{depsdisc} are all in progress, and that is where
any of this would have to land.

\paragraph{What would have to change.} Three things, none of them a competing
standard. All three fit the specification's own extension path, which is why
the ask is a profile rather than a fork. The working group should settle
identity and precedence in 1.1.0,
with the collision rate as the input the original decision lacked. Client
vendors should converge the namespacing they have each already built rather
than each shipping a private one. Registries and CI tooling should report
conformance, the cheapest lever on adoption: it stays near zero while no host
requires it, no tool reports it, and a conformant manifest tells a host no more
than a non-conformant one. An individual author controls none of this and has
to work inside it---\S\ref{sec:takeaways} collects the moves that remain.

\section{Takeaways for Practitioners}
\label{sec:takeaways}

\begin{itemize}[leftmargin=1.2em,itemsep=1.5pt,topsep=2pt]
\item \textbf{Ship the manifest at the bundle root, and declare
      \texttt{\$schema}.} Omitting it is \emph{fatal}---a conforming client
      must refuse to load the plugin---and it is the single reason
      \PctRejected\% of the corpus would be rejected today. Adding it makes
      \PctLoadsFixed\% loadable, and it is the only part of portability fully
      under an author's control.
\item \textbf{Know which of your fields the host will silently drop.} An
      unknown top-level field is non-fatal: the client must ignore it and keep
      loading. \PctLossy\% of bundles would load with at least one field
      discarded, so a plugin that ``works'' can still have had its author's
      intent thrown away with nothing reporting it.
\item \textbf{Put client-specific data in \texttt{extensions}.} The
      specification already sanctions a reverse-domain namespace for it. Only
      \PctExtensions\% of authors use it, while \PctAnyForbidden\% invent a
      top-level field instead and invalidate their manifest for no gain.
\item \textbf{Namespace your capability names now.} Do not name a skill
      \texttt{code-review}. Prefix it. You cannot control what else a user
      installs, and no host will resolve the clash.
\item \textbf{If you must ship duplicate manifests, generate them.}
      \PctDrifted\% of multi-manifest bundles have already drifted and
      \PctDriftName\% disagree about their own name.
\item \textbf{Validate at submission if you run a registry, and test plugin
      \emph{sets}, not plugins.} Gating is the only mechanism here demonstrated
      to produce conformance, and the interesting failures are interactions
      today's tooling never surfaces.
\end{itemize}

\section{Limitations}

We measure manifests committed to a default branch, so a plugin distributed
through an unobserved packaging step is invisible. Capability names come from
directory structure, so the collision analysis covers the \NCapBundles bundles
exposing a named capability rather than the whole corpus. Two plugins sharing a
name could be intentional forks, though the concentration on generic task names
argues against that. Our checker is marginally strict: an open specification
issue notes two cases where the published schema exceeds the normative prose,
shifting the absolute level slightly but no comparison.

Repository dates proxy manifest authorship, which is why \S\ref{sec:temporal}
treats the activity-based cut as primary. Our detector uses the
specification's path conventions plus observed vendor layouts, so an unknown
layout is missed and standalone MCP configurations are
under-counted---no conformance claim rests on that stratum. The corpus is
public GitHub only, and while \S\ref{sec:bias} addresses sampling bias, no
union of keyword searches is a uniform random sample. For practitioner talks
the available text is a third-party summary, so we quote only the speaker's
published words.

\section{Conclusion}

An open format for agent plugins was a necessary step, and it solved a real
problem: a bundle now has one agreed shape. Our measurement of \NBundles
bundles shows adoption is real but early---\PctConformant\% of the corpus
conforms, rising to \ConfRecent\% among actively maintained projects---and that
the residual failure is instructive. Authors keep reaching past the format
toward declaration, dependency, and configuration, while \PctBundlesExposed\%
of capability-exporting bundles sit on a name some other plugin also claims,
with no rule to settle it. Packaging was the visible problem, and the ecosystem
is slowly fixing it. Composition is the one that will break users' agents. A
format standardises how an extension is shipped. A model standardises what
happens when several run at once, which is precisely the problem the ecosystem
now has to solve.

\bibliographystyle{ACM-Reference-Format}
\bibliography{refs}

\balance
\end{document}

%% file: numbers.tex
\newcommand{\TopNameOne}{code-reviewer\xspace}
\newcommand{\TopNameOneOwners}{300\xspace}
\newcommand{\TopNameTwo}{systematic-debugging\xspace}
\newcommand{\TopNameTwoOwners}{221\xspace}

\newcommand{\PctLoads}{6.4\xspace}
\newcommand{\PctRejected}{93.6\xspace}

\newcommand{\PctLoadsFixed}{96.6\xspace}
\newcommand{\NLossy}{27,291\xspace}
\newcommand{\PctLossy}{40.2\xspace}
\newcommand{\NDeepFatal}{2,343\xspace}
\newcommand{\PctDeepFatal}{3.4\xspace}
\newcommand{\NRepos}{30,655\xspace}
\newcommand{\NOwners}{22,901\xspace}
\newcommand{\NBundles}{68,072\xspace}
\newcommand{\NSkillsStandalone}{157,062\xspace}
\newcommand{\NMcpStandalone}{1,379\xspace}
\newcommand{\NCaps}{409,935\xspace}
\newcommand{\NLocated}{13,417\xspace}
\newcommand{\PctLocated}{19.7\xspace}
\newcommand{\NConformant}{4,237\xspace}
\newcommand{\PctConformant}{6.2\xspace}

\newcommand{\PctDeclareComponents}{32\xspace}
\newcommand{\PctAnyForbidden}{43\xspace}
\newcommand{\PctBundlesExposed}{81\xspace}
\newcommand{\PctBundlesExposedXOwner}{74\xspace}
\newcommand{\NCapBundles}{31,901\xspace}
\newcommand{\PctSingleBundleRepos}{84\xspace}
\newcommand{\PctTopRepoShare}{1.7\xspace}

\newcommand{\NSkillNames}{94,837\xspace}
\newcommand{\PctSkillNamesShared}{28\xspace}
\newcommand{\PctSkillInstContested}{78\xspace}
\newcommand{\MaxSkillClaims}{1,012\xspace}
\newcommand{\ConfPre}{3.2\xspace}
\newcommand{\ConfPost}{19.9\xspace}
\newcommand{\NPre}{55,881\xspace}
\newcommand{\NPost}{12,191\xspace}
\newcommand{\NLive}{45,796\xspace}
\newcommand{\ConfLive}{9.3\xspace}
\newcommand{\NRecent}{38,331\xspace}
\newcommand{\ConfRecent}{10.4\xspace}
\newcommand{\SchemaRecent}{15.5\xspace}
\newcommand{\NNeutralBundles}{36,157\xspace}
\newcommand{\SpecDate}{24 July 2026\xspace}
\newcommand{\CorpusName}{AgentPluginZoo\xspace}
\newcommand{\NNeutralOnly}{10,511\xspace}
\newcommand{\ConfNeutralOnly}{23.2\xspace}
\newcommand{\NOneVendor}{49,309\xspace}
\newcommand{\ConfOneVendor}{2.0\xspace}
\newcommand{\NMultiVendor}{8,252\xspace}
\newcommand{\ConfMultiVendor}{9.7\xspace}

\newcommand{\PctExtensions}{1.9\xspace}
\newcommand{\NExtNamespaces}{129\xspace}
\newcommand{\NMultiManifest}{9,857\xspace}
\newcommand{\PctMultiManifest}{14.5\xspace}
\newcommand{\NDriftSampled}{3,998\xspace}
\newcommand{\PctDrifted}{95.6\xspace}
\newcommand{\PctDriftName}{7.3\xspace}

%% file: ledger.tex
recursive tree expansion & --- & --- & 226,513 \\
code search: .claude-plugin/plugin.json & 80 & 24,589 & --- \\
code search: marketplace indexes & 88 & 2,265 & --- \\
code search: .codex-plugin/plugin.json & 10 & 1,780 & --- \\
code search: explicit v1.0.0 \$schema & 13 & 1,703 & --- \\
code search: .github/plugin/plugin.json & 1 & 93 & ---